\documentclass[%
 reprint,
 amsmath,amssymb,
 aps,
]{revtex4-2}
\usepackage{graphicx}% Include figure files
\usepackage{dcolumn}% Align table columns on decimal point
\usepackage{bm}% bold math

\makeatletter
\def\paragraph{\@startsection
    {paragraph}    % name
    {4}            % level
    {\z@}          % indent
    {1ex \@plus1ex \@minus.2ex}  % beforeskip
    {0pt}          % afterskip
    {\normalfont\normalsize\itshape}} % style
\makeatother

\begin{document}

\title{
Metropolitan quantum key distribution using a GaN-based room-temperature telecommunication single-photon source }

\author{Haoran Zhang$^1$}
\author{Xingjian Zhang$^2$}
\author{John Eng$^1$}
\author{Max Meunier$^1$}
\author{Yuzhe Yang$^1$}
\author{Alexander Ling$^{2,3,}$}
\email{alexander.ling@nus.edu.sg}
\author{Jesús Zúñiga-Pérez$^{1,4,}$}
\email{jesus.zuniga@ntu.edu.sg}
\author{Weibo Gao$^{1,2,5,}$}
\email{wbgao@ntu.edu.sg}
\affiliation{$^1$Division of Physics and Applied Physics, School of Physical and Mathematical Sciences, Nanyang Technological University, Singapore, Singapore}
\affiliation{$^2$Centre for Quantum Technologies, National University of Singapore, Singapore, Singapore}
\affiliation{$^3$Department of Physics, National University of Singapore, Singapore, Singapore}
\affiliation{$^4$MajuLab, International Research Laboratory IRL 3654, CNRS, Université Côte d’Azur,
Sorbonne Université, National University of Singapore, Nanyang Technological University,
Singapore, Singapore}
\affiliation{$^5$The Photonics Institute and Centre for Disruptive Photonic Technologies, Nanyang Technological University, Singapore, Singapore}

\begin{abstract}
Single-photon sources (SPS) hold the potential to enhance the performance of quantum key distribution (QKD). QKD systems using SPS often require cryogenic cooling, while recent QKD attempts using SPS operating at room-temperature have failed to achieve long-distance transmission due to the SPS not operating at telecommunication wavelength. In this work, we have successfully demonstrated QKD using a room-temperature SPS at telecommunication wavelength. The SPS used in this work is based on point defects hosted by gallium nitride (GaN) thin films grown on sapphire substrates. We employed a time-bin and phase encoding scheme to perform the BB84 and reference-frame-independent QKD protocols over a 33 km fiber spool, achieving a secure key rate of $7.58\times 10^{-7}$ per pulse. Moreover, we also implemented a metropolitan QKD experiment over a 30 km deployed fiber, achieving a secure key rate of $6.06\times 10^{-8}$ per pulse. These results broaden the prospects for future use of SPS in commercial QKD applications.
\end{abstract}

\maketitle

\paragraph*{Introduction}
---Single photons are fundamental resources in quantum optics research and quantum technologies. Advances in single-photon sources (SPS) \cite{aharonovich2016solid,senellart2017high} have opened new opportunities in quantum communication \cite{beveratos2002single,nilsson2013quantum}, quantum computing \cite{northup2014quantum,ding2023high} and quantum sensing \cite{gao2015coherent}. In the context of quantum key distribution (QKD), an ideal SPS can approach the channel loss limit performance. Additionally, a QKD system utilizing an SPS can eliminate the need for decoy states, thereby reducing experimental and data processing complexity. Most importantly, SPS-based QKD systems should surpass the coherent-state limit \cite{zhang2024experimental}, which highlights the potential of single photons in enhancing QKD performance.

Many studies on QKD with SPS have been reported in recent years \cite{vajner2022quantum,heindel2023quantum}. However, most of these sources \cite{kupko2020tools,gao2022quantum,zahidy2024quantum,yang2024high} require cryogenic cooling, which hinders their widespread commercial application. While approaches using room-temperature SPS in hexagonal boron nitride (hBN) \cite{al2023quantum,zeng2022integrated,tran2016quantum} have addressed this issue, unfortunately these sources do not operate at telecommunication wavelengths, thereby limiting their applications for long-distance fiber-based QKD.

In 2018, a solid-state SPS emitting in the telecommunication wavelength range and operating at room-temperature was discovered in gallium nitride (GaN) crystals \cite{zhou2018room}. These SPS display record-high brightness for a point-defect emitter, stable triggered photoluminescence (PL), short lifetime in the order of hundreds of picoseconds, relatively narrow linewidth, and high purity \cite{meunier2023telecom}. These characteristics make it an ideal source candidate for realizing practical long-distance QKD. Furthermore, GaN technology is industrially mature and optoelectronic and electronic devices based on this technology \cite{zuniga2016polarity,feezell2018invention,meneghini2021gan} are in widespread use. Thus, one can expect these sources to be combined, within the same chip, with devices such as light-emitting diodes or lasers and transistors, enabling the development of compact and monolithic room-temperature SPS.

\begin{figure}[t]
    \centering
    \includegraphics[width=1\linewidth]{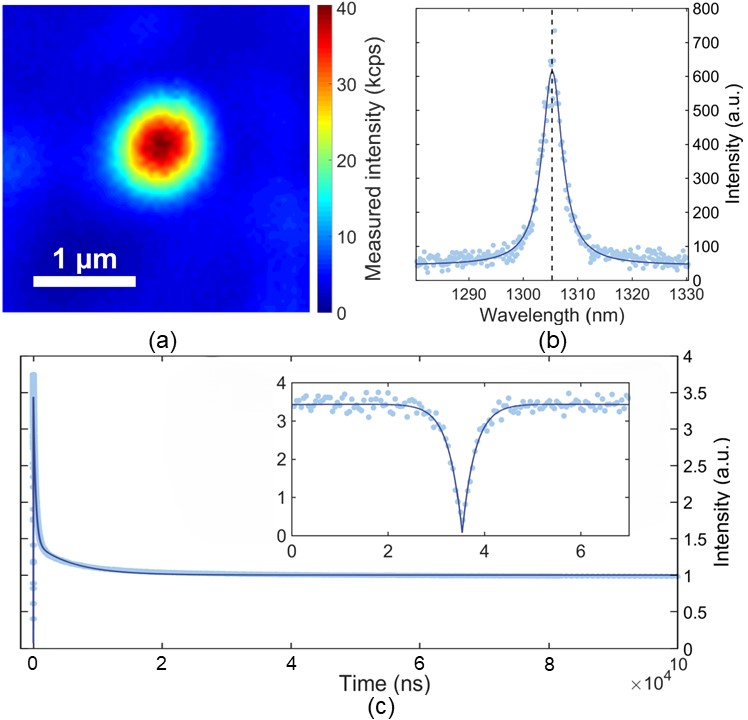}
    \caption{GaN-based SPS. (a) Photoluminescence mapping of SPS. (b) Spectrum of the SPS.  (c) Normalized second-order correlation function $g^{(2)}(\tau)$ using a continuous-wave laser pump. }
\end{figure}

In this article we report a proof-of-principle experiment realizing a QKD system using a GaN-based room-temperature SPS at telecommunication wavelength (O-band). The QKD system is tested both locally and remotely. The metropolitan QKD experiment is conducted between the sites of Nanyang Technological University (NTU) and National University of Singapore (NUS), both in Singapore, using a fiber link deployed between the 30 km distant locations. By using a time-bin and phase encoding scheme instead of a polarization encoding scheme, the quantum bit error rate (QBER) limited by the large polarization mode dispersion (PMD) \cite{galtarossa2005polarization} in the deployed fiber is avoided. The key rate is calculated and the performance analyzed with different block sizes. The entire experiment demonstrates that GaN-based room-temperature SPS can be effectively used for metropolitan QKD.

\paragraph*{Preparation of single-photon source}

\begin{figure*}[htbp]
    \centering
    \includegraphics[width=1\linewidth]{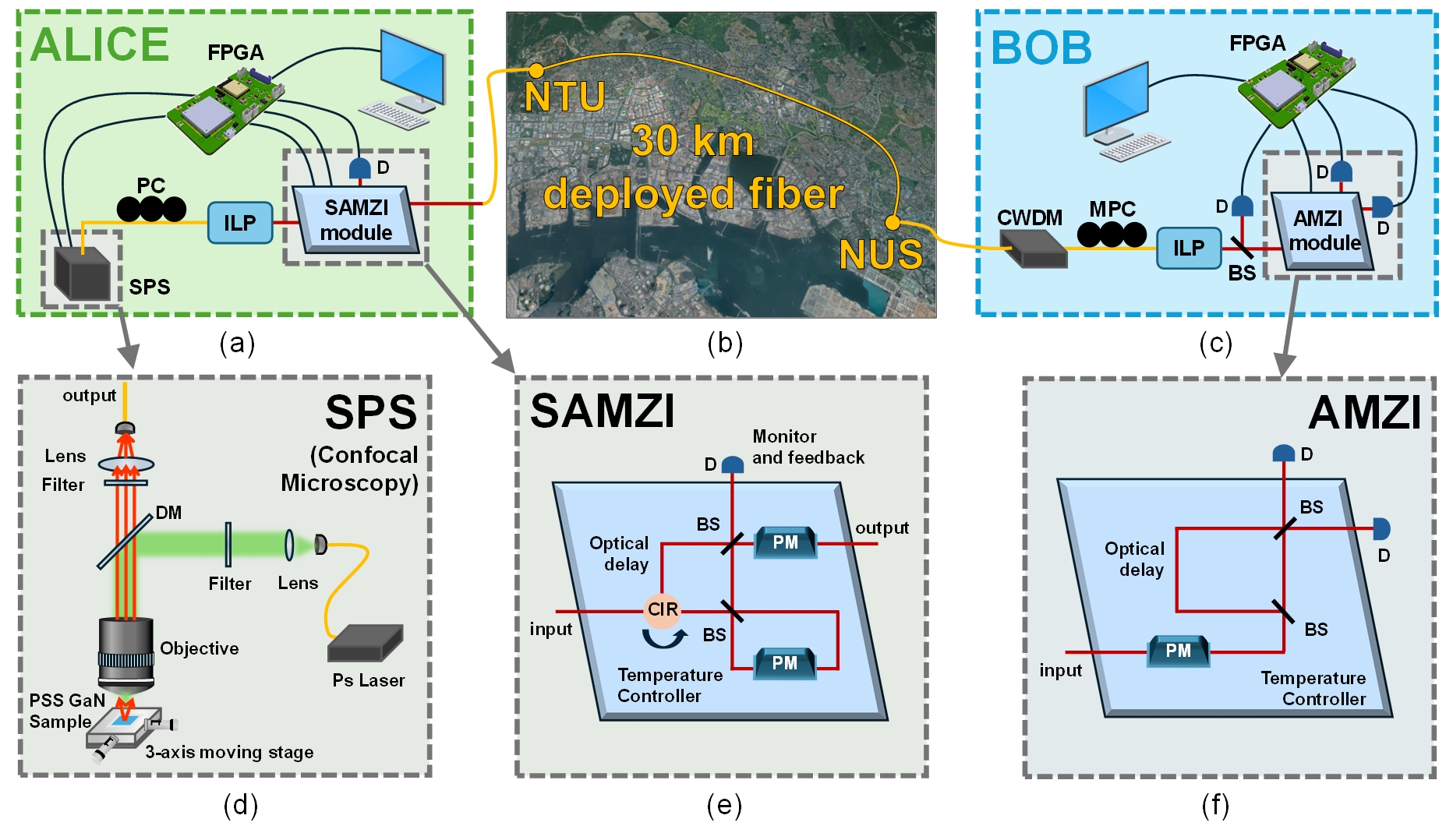}
    \caption{Experimental setup of QKD system. DM: dichroic mirror; SPS: single-photon source; PC: polarization controller; ILP: in-line polarizer; MPC: motorized polarization controller; D: superconducting nanowire single-photon detector; FPGA: field programmable gate array and time tagger; CWDM: coarse wavelength division multiplexing filter; CIR: circulator; BS: polarization maintaining beam splitter; PM: electro-optic phase modulator; Black line: cable; Orange line: single-mode fiber; Red line: polarization maintaining fiber. (a) Schematic of the QKD setup at Alice's end. (b) Schematic of the layout of the 30 km deployed fiber (underground).  (c) Schematic of the QKD setup at Alice's end. (d) Confocal setup for exciting the SPS and collecting the single-photons. (e) Illustration of SAMZI module. SAMZI: Sagnac asymmetric Mach-Zehnder interferometer. (f) Illustration of AMZI module. AMZI: asymmetric Mach-Zehnder interferometer.}
    \label{fig:setup}
\end{figure*}

---The optical excitation of the GaN-based SPS and the collection of the emitted single-photons are achieved using a homemade confocal microscopy setup. To enhance the PL count rate of the SPS, the GaN thin film employed in this study was grown on a patterned sapphire substrate, which improves the extraction efficiency. To use in a QKD system, the emission of the SPS must be synchronized with modulation and measurement components. Therefore, a synchronized pulsed laser operating at 80 MHz is used to excite the defect. Figure 1(a) shows the PL mapping results around the chosen defect, with a measured count rate of $\sim 40 $ kcps when excited by an 80 MHz repetition rate pulsed laser. The excitation laser power was set near saturation at 0.17 mW. Figure 1(b) presents the spectrum of the SPS, revealing a center wavelength of 1305.4 nm and a full width at half maximum (FWHM) of $8.96\pm 0.13$ nm, when fitted with a Lorentzian function. This wavelength is very close to 1310 nm, aligning well with the operating wavelength of commercial optical communication devices. Due to interaction with phonons, which are enhanced at room-temperature, and local time-varying fluctuations (e.g. charged defects), the actual spectrum shape is not a pure Lorentzian \cite{borri2001ultralong}.  Figure 1(c) shows the normalized second-order correlation function $ g^{(2)}(\tau) $ using a 0.1 mW continuous-wave laser pump. Considering the total time jitter of the entire setup, which is comparable to the lifetime of the SPS, the measured results are expected to be strongly affected by the instrument response function, which needs to be deconvoluted \cite{fishman2023photon}. Similar to many solid-state defects, the $g^{(2)}(\tau) $ function needs to account for the presence of additional energy levels besides those responsible for the single-photon emission \cite{meunier2023telecom}. For a 4-level model:
\begin{equation}
     g^{(2)}(\tau) = 1 - Ae^{-|\tau|/\tau_1} + Be^{-|\tau|/\tau_2}+ Ce^{-|\tau|/\tau_3} 
\end{equation}
where $ \tau_1 $, $ \tau_2 $ and $ \tau_3 $ are primarily determined by the excitation and decay rates between the ground, excited and metastable states.  The fitted value at $ g^{(2)}(0) = 1 - A + B+C $ is $0.065 \pm 0.061$. 

\begin{figure}[htbp]
    \centering
    \includegraphics[width=1\linewidth]{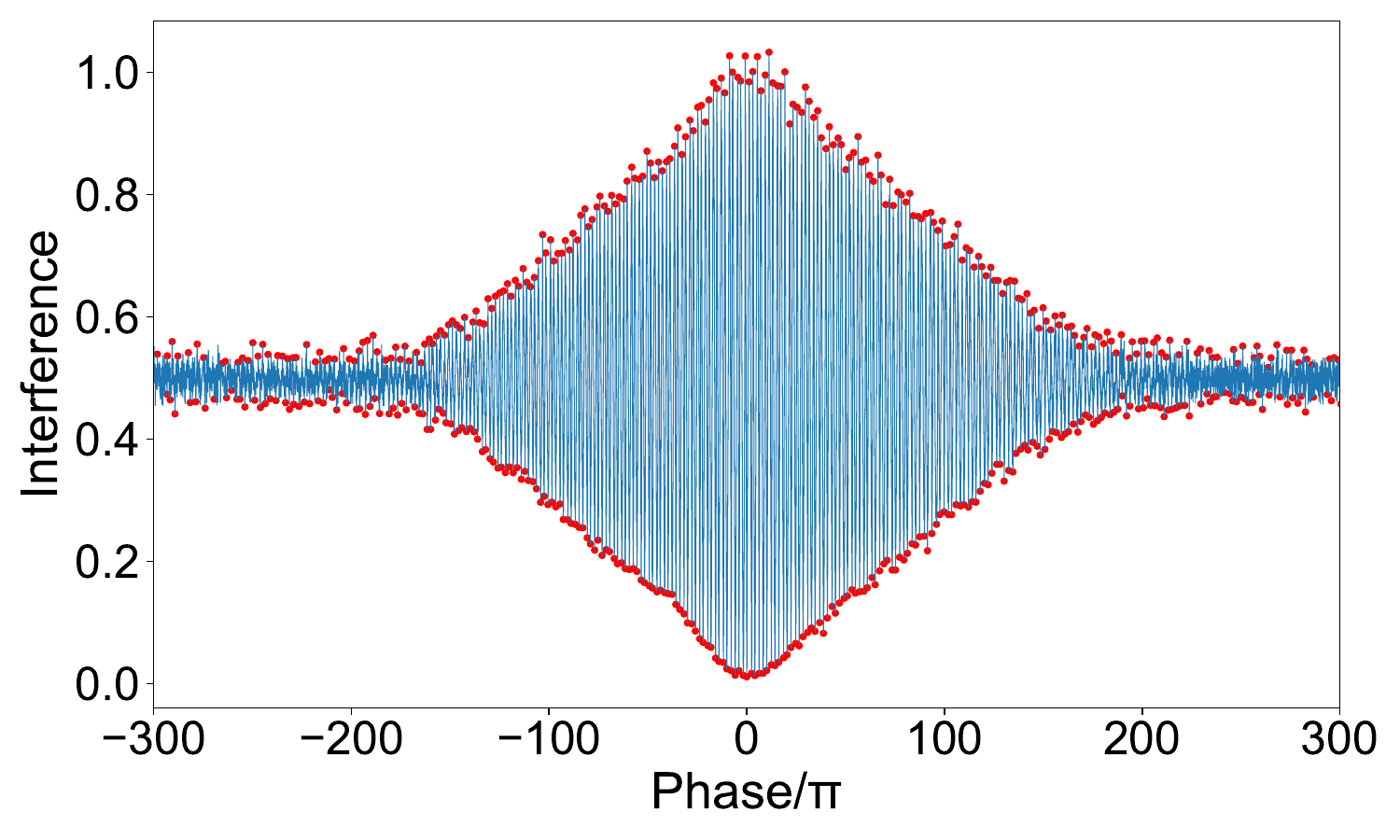}
    \label{fig:Coherent}
    \caption{Experimental results showing the variation in interference intensity with phase difference. Each data point represents a 1-minute accumulation, normalized against the intensity of non-interference pulses.}
\end{figure}

\paragraph*{Experimental setup of QKD system}
---Given the wide linewidth of the SPS at room temperature and the high PMD of approximately $1.50 ps$ ($2.71ps/\sqrt{km}$) across the entire commercial fiber link between NTU and NUS, time-bin and phase-encoded photons are preferred over polarization-encoded photons to achieve a stable and low QBER in metropolitan QKD experiments, as they are inherently free from birefringence compensation requirements. Figure 2 shows the experimental setup of the entire QKD system. As shown in Figure 2(a) and Figure 2(c), the whole QKD system is synchronized by two FPGA boards operating at 80 MHz. The CWDM filter is placed at Bob's end to filter out high background photons in the commercial fiber. The SAMZI module is used to modulate time-bin and phase states at Alice's end. With an identical optical delay compared to the SAMZI module, the AMZI module at Bob's end demodulates the phase states generated at Alice's end. For measuring time-bin states, the superconducting nanowire single-photon detector (SNSPD) directly detects the arrival time of photons. Figure 2(b) is a schematic of the layout of NTU and NUS labs, connected by a 30 km commercially deployed fiber with 15.52 dB loss. Figure 2(d) illustrates the light source employed in the QKD system. The employed confocal microscopy geometry ensures that the excitation laser is focused at the same point from which the collected photons originate, within the GaN thin film. Figure 2(e) and Figure 2(f) show the details of the SAMZI and AMZI modules. In the SAMZI module, the input photon first enters the Sagnac loop and then interferes, splitting into the long and short paths. The PM in the Sagnac loop generates two time-bin states or a phase state by adjusting the probability of the paths the photon takes. Specifically, the SAMZI module produces time-bin states when the modulation phase of this PM is set to $0$ or $\pi$, and phase states when the modulation phase is set to $\pi /2$. Compared to modulating time-bin states using an electro-optic intensity modulator, the Sagnac structure is more robust because it eliminates the need for bias voltage compensation. Another PM modulates the phase states by adjusting the relative phase of two sub-pulses. The other output of the SAMZI is detected to monitor the SPS's operation and to provide feedback to compensate for mechanical vibrations from the environment. In the AMZI module, a PM adjusts the measurement basis of phase states. Based on the interference output, the relative phase of the pulses passing through the short-long and long-short paths of the two interferometers is measured.

\paragraph*{Interference of Single-photon Source}
---The QBER of phase states increases if the interference visibility of the SPS diminishes. To minimize this effect, each interferometer is equipped with a temperature controller to adjust and stabilize the optical delay of the two interferometers. By fixing the temperature of one interferometer and continuously adjusting the temperature of the second one, an optimal temperature maximizing the interference visibility of the SPS to $98\%$ can be found. As shown in Figure 3, the interference intensity oscillates continuously, allowing for the counting of all interfered peaks. The interference visibility decreases to $0.5$ when the phase difference reaches $88 \pi$ ($\sim 57.4 \mu m$). This highlights the importance of using temperature controllers. Additionally, maintaining a stable temperature for the interferometers can reduce or eliminate phase drift, simplifying the compensation process.

\paragraph*{Experimental results and simulation of QKD system}

---The entire setup enables QKD protocols to be implemented with time-bin and phase states. We define the time-bin states as the states in the 
 $Z$ basis, $|0\rangle$ and $|1\rangle$, while the phase states are defined as the states in $X$ or $Y$ basis, $(|0\rangle \pm |1\rangle)/\sqrt{2}$ or $(|0\rangle \pm i|1\rangle)/\sqrt{2}$. The time-bin states are used for generating raw key while the phase states are used for monitoring the quantum channel \cite{boaron2018secure}, which is suitable for performing both the BB84 protocol \cite{bennett2014quantum} and the reference-frame-independent QKD (RFI-QKD) protocol \cite{laing2010reference}. For RFI-QKD, time-bin states are still used for generating raw key but the phase states should be prepared and measured in both the $X$ and $Y$ basis.

To comprehensively show the performance of the QKD system, we tested these two protocols both locally and remotely. For the local test, the preparation and measurement were both performed at NTU lab using a 33 km fiber spool as the quantum channel. Additionally, a 0 km test, where Alice and Bob are connected directly, was conducted to obtain the baseline performance of the QKD system. For the remote test, the measurement was performed at NUS lab. The test results of QBER are shown in Table 1 (also see in supplementary material). The QBER of the two protocols using the four different quantum channels were obtained. $E_z$ and $E_x$ represent the QBER of time-bin states in the $Z$ basis and the QBER of phase states in $X$ basis, respectively. Parameters $Q$ and $R$, which will be elaborated on below and supplementary material, represent the gain and the secure key rate per pulse, respectively. The deployed fiber has a measured background rate around 1 kcps, attributed to multiple fiber switches within a network that also carries conventional communications traffic. The QBER after background correction was calculated for showing the misalignment of quantum signals. It is interesting to note that the background noise does not significantly influence the secure key rate if vacuum decoy states are added \cite{ma2005practical}.
\begin{table*}[t]
    \centering
    \begin{tabular}{|m{2.9cm}|c |c| c| c| c| c |c|c |c|}
    \hline
        Parameter & Distance &Loss & $E_z$ & $E_x$ & $Q_z$ & $Q_x$ & $Q_c$ & $R_{BB84}$ & $R_{RFI}$\\
        \hline
        Local & 0 km &0 dB& $0.89\%$ &$1.88\%$& $1.63\times 10^{-5}$ & $4.63\times 10^{-6}$& $4.51\times 10^{-6}$ & $1.28\times 10^{-5}$& $1.34\times 10^{-5}$ \\
        \hline
        Local & 33 km &10.44 dB& $1.78\%$ &$4.71\%$& $1.03\times 10^{-6}$ & $3.54\times 10^{-7}$& $3.50\times 10^{-7}$ & $7.58\times 10^{-7}$& $8.50\times 10^{-7}$ \\
        \hline
        Metropolitan & 30 km &15.52 dB& $4.16\%$ &$10.2\%$& $2.32\times 10^{-7}$ & $8.11\times 10^{-8}$& $6.06\times 10^{-8}$ & $5.60\times 10^{-8}$& $6.07\times 10^{-8}$ \\
        \hline
       Metropolitan (background corrected) & 30 km &15.52 dB& $2.19\%$ &$5.18\%$& $2.23\times 10^{-7}$ & $7.28\times 10^{-8}$ & $5.30\times 10^{-8}$ & $1.18\times 10^{-7}$ & $1.18\times 10^{-7}$\\
        \hline
    \end{tabular}
    \caption{Experimental results of QKD.}
    \label{tab:keyrate}
\end{table*}

The secure key rate of the above two QKD protocols using a SPS is given by \cite{cai2009finite,liang2014proof}:
\begin{equation}
\begin{split}
    R=Q_z [ A_z(1-I_E)-f_{EC} \cdot h(E_z)-7\sqrt{\frac{\log_2(2/\hat{\epsilon})}{n_z}} \\ -\frac{2}{n_z}\log_2 \frac{1}{\epsilon_{pa}}- \frac{1}{n_z}\log_2 \frac{2}{\epsilon_{cor}}]
    \end{split}
\end{equation}
where $h(x)$ is binary Shannon entropy, $f_{EC}$ is the error correction efficiency, $Q_z$ is the gain of time-bin states in $Z$ basis, and $n_z$ is the total count of time-bin states for the generating key. $I_E$ is the leaking information of single photon, $\epsilon_{pa}$ and $\hat{\epsilon} $ represent the security parameter in privacy amplification and smooth min-entropy, respectively, and $\epsilon_{cor}$ indicates the failure probability of the correctness \cite{tomamichel2012tight}. The parameter $A_z=\frac{Q_z-P_m}{Q_z}$ is the ratio of single photon detection in $Z$ basis, where $P_m$ is the probability of multi-photon emission at the Alice's end. For SPS, we have $P_m \leq \mu ^2 g^{(2)}_{pul}(0)/2$, where the output mean photon number $\mu= 8.955\times 10^{-5}$, pulsed second-order correlation $g^{(2)}_{pul}(0)=0.356 \pm 0.007$. 

The experimental results of secure key rate are shown in Table 1, where error correction efficiency is set to $f_{EC}=1.1$. The secure key rates are calculated assuming infinite length, indicating that the metropolitan QKD system can work. However, the finite-size effect should also be analyzed to evaluate the appropriate operating time. As shown in Figure 4 and supplementary material, the secure key rate using this setup is simulated under experimental parameters obtained from the tests. The block size is set to \(N \rightarrow \infty\), \(N = 10^{10}\), \(N = 10^{12}\), and \(N = 10^{14}\), corresponding to working times from several minutes to more than one week if the system operates at 80 MHz. For RFI-QKD, the finite-size problem is addressed by the free-running RFI-QKD protocol \cite{tang2022free}, where data slices with similar misalignment angles \(\theta\) are combined into 12 groups. The parameters for the finite-size effect \cite{scarani2008security,sheridan2010finite} are provided in detail in the supplementary material. According to the 0 km test results, the overall detection efficiencies of time-bin and phase states are $\eta_z=0.182$ and $\eta_x=0.0784$, respectively. The time-bin states and phase states are supposed to have intrinsic QBER of $e_z=0.89\%$ and $e_x=1.88\%$. The dark count rate of one measurement is set to $2\times 10^{-8}$ per channel use. The simulation results suggest that a suitable block size $N$ for long-distance QKD should be at least $10^{10}$, corresponding to an operating time of several minutes.

\begin{figure}[htbp]
    \includegraphics[width=1\linewidth]{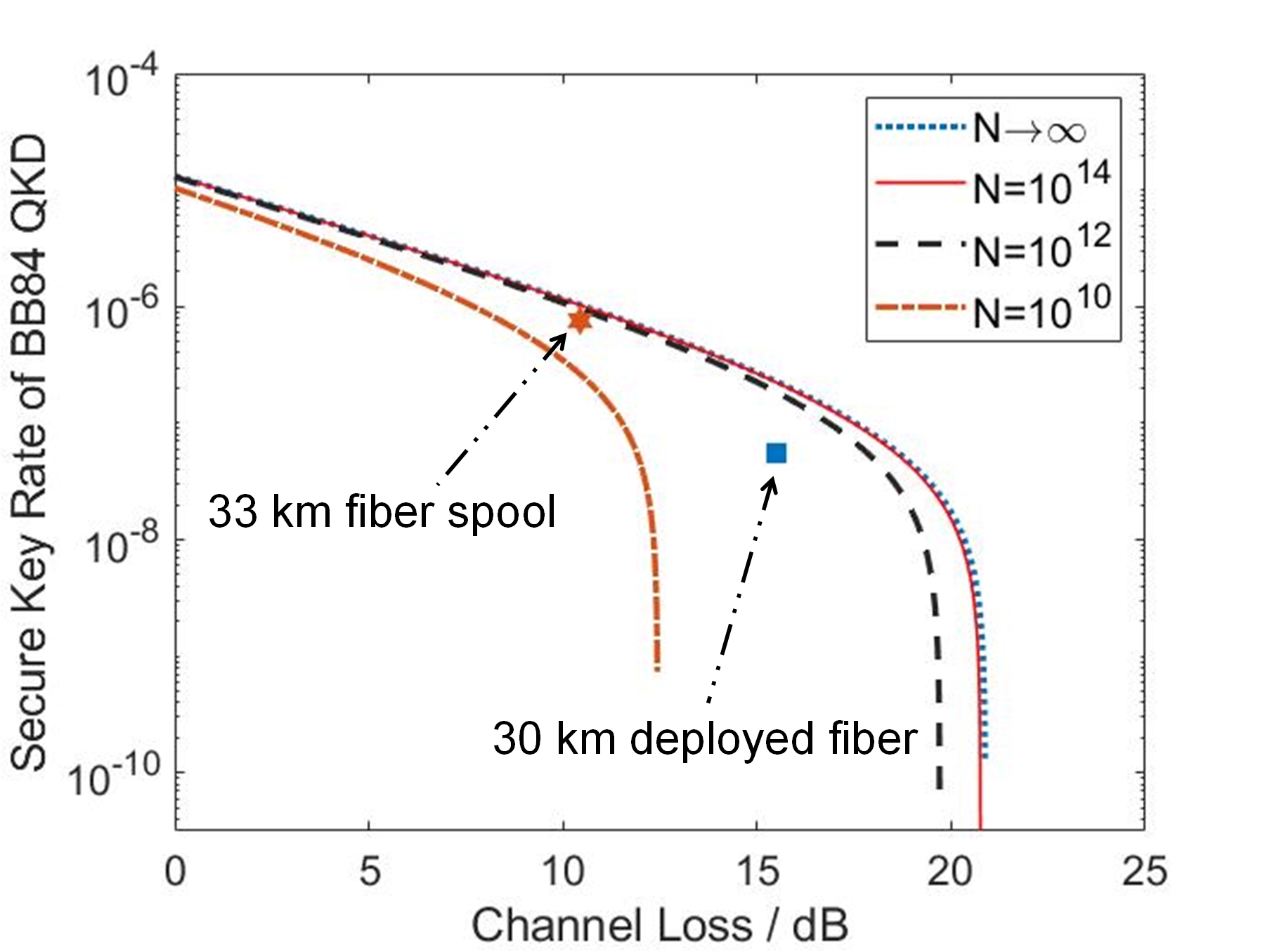}
    \centering
    \label{fig:keyrate}
\caption{Simulation of secure key rate varies with channel loss for BB84 protocol using experimental parameters.}
\end{figure}

\paragraph*{Discussion}
---In our metropolitan experiment, we also tested polarization encoding. The PMD of the deployed fiber resulted in varying increases in the QBER depending on the polarization state basis. Although the intrinsic QBER due to the preparation and measurement setup was less than $1\%$, we observed QBER of $16.7\%$ and $35.1\%$ (reduced to $11.0\%$ and $30.2\%$ with background correction) for two polarization basis. These results indicate that more demanding efforts, such as further narrowing the bandwidth of the SPS or effectively compensating for the PMD effect, are necessary to realize polarization encoding schemes over our metropolitan fiber link.

In contrast, using time-bin and phase encoding reduced the QBER to approximately $4\%$ and $10\%$, respectively ($2\%$ and $5\%$ with background correction). The QBER for phase encoding was higher than that for time-bin encoding. Approximately half of this increase was due to background noise, as the additional attenuation introduced by the AMZI module lowered the signal-to-noise ratio. The remaining QBER increase after background correction might be attributed to challenges in phase drift compensation. Due to the relatively low count rate, accurate compensation was difficult, but the QBER could be further reduced by increasing the repetition rate and improving temperature control stability.

Alternatively, using a free-running RFI-QKD protocol avoids the issue of phase drift compensation. However, data accumulation and finite size effects could still be affected by fluctuations in the interferometers. Thus, precise temperature control \cite{zhang2021polarization} is crucial for enhancing the performance of both BB84 and RFI-QKD protocols.

\paragraph*{Conclusion}
   
---Overall, we report the first proof-of-principle implementation of a QKD system using a room-temperature telecommunication wavelength SPS and further test its feasibility and applicability in a metropolitan experiment.  
To enhance the performance of the current QKD setup, it is crucial to reduce the noise of the deployed commercial fiber but also to improve the quality of the SPS, particularly its brightness. To do so one can think of increasing its internal quantum efficiency \cite{hoang2015ultrafast} as well as the collection efficiency \cite{sapienza2015nanoscale} through nanofabrication of photonic structures around the SPS \cite{meunier2023telecom}. Furthermore, assessing the nature of the defect at the origin of the SPS \cite{luo2024room,geng2023optical,meunier2023telecom} would certainly contribute to their controlled fabrication and integration into photonic structures, enabling the production of better-performance SPS. Additionally, due to the mature GaN technology, in particular that associated to GaN-based light sources, GaN SPS have the potential for being electrically-injected if the emitting defects can be introduced in a controlled manner into p-i-n stacks \cite{deshpande2013electrically}, which in turn could be integrated into cavities or waveguides \cite{vico2012high,somaschi2016near}. Consequently, GaN telecommunication SPS could be the ideal choice for commercial QKD applications if their potential advantages are fully realized.

\begin{acknowledgments}
This work is supported by the National Research Foundation, Singapore and A*STAR under its Quantum Engineering Programme (NRF2021-QEP2-01-P02, National Quantum-Safe Network, NRF2021-QEP2-04-P01, NRF2022-QEP2-02-P13), ASTAR (M21K2c0116, M24M8b0004), Singapore National Research foundation (NRF-CRP22-2019-0004, NRF2023-ITC004-001, NRF-CRP30-2023-0003, NRF-MSG-2023-0002), Singapore Ministry of Education Tier 2 Grant (MOE-T2EP50222-0018), by the French government through the National Research Agency (ANR) in the context of the Plan France 2030 through project reference ANR-22-PETQ-0011, and by the ANR through project ANR-22-CE47-0006-01. We acknowledge Netlink Trust for the provisioning of the fibre network. We thank the support from Dieter Schwarz Stiftung gGmbH.

\end{acknowledgments}

\bibliography{ref}

\end{document}